\documentclass[floats,floatfix,amssymb,prd,twocolumn,superscriptaddress,nofootinbib]{revtex4-2}

\usepackage[english]{babel}
\usepackage[utf8]{inputenc}
\usepackage{amsmath}
\usepackage{mathbbol}
\usepackage{amssymb}
\usepackage{bbold}
\usepackage{graphicx,amsfonts}
\usepackage{epsfig}
\usepackage[svgnames]{xcolor}
\definecolor{crimsonglory}{rgb}{0.75,0.0,0.2}
\usepackage[colorlinks=true,
linkcolor=crimsonglory,
urlcolor=crimsonglory,
citecolor=crimsonglory]{hyperref}
\usepackage{bm}
\usepackage{mathrsfs}
\usepackage{mathtools}
\usepackage{enumerate}
\usepackage{amsthm}
\usepackage{bbm}
\usepackage{comment}
\usepackage{physics}
\usepackage{url}
\usepackage{upgreek}
\usepackage[title]{appendix}
\usepackage{float} 
\usepackage{rotating}
\usepackage{booktabs} 

\begin{document}

\title{Vortices without inflow: bound spectra in horizonless rotational analogs}
%
\author{H. S. Vieira}
\affiliation{Theoretical Astrophysics, Institute for Astronomy and Astrophysics, University of T\"{u}bingen, 72076 T\"{u}bingen, Germany}
\author{Kyriakos Destounis}
\affiliation{Theoretical Astrophysics, Institute for Astronomy and Astrophysics, University of T\"{u}bingen, 72076 T\"{u}bingen, Germany}
\affiliation{CENTRA, Departamento de Física, Instituto Superior Técnico – IST, Universidade
de Lisboa – UL, Avenida Rovisco Pais 1, 1049-001 Lisboa, Portugal}
%
\begin{abstract}
Analog gravity experiments are making remarkable strides in unveiling both the classical and quantum nature of black holes. By harnessing diverse states of matter, contemporary tabletop setups now replicate strong-field phenomena typically confined to the enigmatic regions surrounding black holes. Through these modern gravity simulators, physical processes once considered elusive may finally be brought into experimental reach.
In this work, we investigate the spectrum of massless scalar excitations propagating within the effective geometry of a rotating acoustic metric. Specifically, we build an analog vortex-like spacetime endowed with a tunable parameter that emulates the geometry of a rotating gravitational background. This model accommodates both the presence of a sonic horizon, characteristic of an acoustic black hole for non-zero tuning parameters, and its absence when the parameter vanishes, yielding a horizonless, purely rotational vortex flow devoid of radial inflow. 
We focus on the case where the vortex flow is purely rotational. The resulting spectral properties is found to be qualitatively consistent with that observed in recent experimental realizations of giant multiply quantum vortices featuring solid or hollow cores. This correspondence suggests that the analog spacetime used here holds significant potential to replicate, qualitatively, the phenomenology of cutting-edge laboratory experiments. In doing so, it offers new insight into the intricate landscape of analog black-hole spectroscopy and, potentially, the resonant topography of bounded, rotating astrophysical environments around black holes.
\end{abstract}

\maketitle

\section{Introduction}\label{Introduction}

Gravitational-wave (GW) events from coalescing binaries are currently flooding ground-based detectors with data from compact binary sources. The LIGO-Virgo-KAGRA (LVK) collaboration has successfully detected more than $200$ GW events \cite{KAGRA:2021vkt}. These confirmed observations are sourced by mostly black hole (BH) and neutron star (NS) mergers, as well as BH-NS binary systems. Their characteristic imprint on the LVK detectors is divided in three stages; an initial \emph{inspiral} stage, where the two compact objects meander around a center of mass whilst emitting GWs, thus their orbital distance is shrinking, followed by a \emph{merger} phase, where the two objects plow into each other violently. The final stage is dubbed the \emph{ringdown}, where the vibrating remnant tends into equilibrium by radiating away its final fluctuations in a series of damped sinusoids, known as the quasinormal modes (QNMs) \cite{Kokkotas:1999bd,Berti:2009kk,Konoplya:2011qq}.

Concurrently, analog gravity simulators have shown great progress in the field of classical and quantum BH phenomenology \cite{Visser:1997ux,Fischer:2001jz,Barcelo:2005fc,Barcelo:2018ynq,Braunstein:2023jpo,Yang:2024fql}. Contemporary technological advances in controlling and manipulating such emulators have enabled the experimental realization of acoustic analogs of gravitational BHs \cite{Unruh:1980cg}. The past decade witnessed a large variety of laboratory experiments testing a diverse set of aspects of BH physics that led to tentative evidence of analog Hawking radiation \cite{Weinfurtner:2010nu,Steinhauer:2014dra,Steinhauer:2015saa,MunozdeNova:2018fxv,Kolobov:2019qfs,Tian:2020bze}, superradiance \cite{Richartz:2014lda,Torres:2016iee} and imprints of QNMs in perturbed analogs \cite{Torres:2019sbr,Torres:2020tzs}. Bridging the gap between analog BH experiments and realistic GWs is now in full bloom. The latest experimental and theoretical developments in these areas illustrate that there is an overabundance of possibilities \cite{Baak:2023zjf}, even at cosmological scales \cite{Krusius:1998ms,Volovik:2003fe,Fedichev:2003bv}. 

Even though analog gravity was initially proposed by Unruh \cite{Unruh:1980cg}, in order to understand the quantum nature of spacetime through analog BHs with sonic horizons, beyond which acoustic excitation become supersonic and cannot escape the hole, at the moment the only observed phenomenon from binary mergers is the emission of GWs, as described above, which is a purely classical manifestation of strong-field gravity. The study of a remnant's QNMs, dubbed as the \emph{BH spectroscopy program} \cite{Dreyer:2003bv,Ota:2021ypb,Destounis:2023ruj,Berti:2025hly}, can serve as a validity test for analog BH spectroscopy \cite{Torres:2019sbr}. Since the experimental results from analog BH laboratories correctly hint the existence of QNMs, and even quasibound states (QBSs) \cite{Patrick:2018orp}, so the observed analog Hawking radiation may provide, at least qualitatively, legitimate evidence for the quantum nature of gravitation \cite{Hawking:1974rv,Hawking:1975vcx}.

Most BH analogs that are built with fluids \cite{Unruh:1994je,Berti:2004ju,Coutant:2016vsf,Torres:2022bto,Vieira:2021xqw,Vieira:2025ljl}, optical lasers \cite{Corley:1998rk,Coutant:2009cu,Leonhardt:2008js,Reyes:2017,Peloquin:2015rnl}, Bose-Einstein condensates \cite{Garay:2000jj,Fedichev:2003id,Gooding:2020scc,Liao:2018avv,Lahav:2009wx,Ribeiro:2021fpk,Ribeiro:2022gln,Baak:2022hum,Fabbri:2020unn,Solnyshkov:2018dgq,Klaers:2010,Fischer:2004bf,Vieira:2023ylz}, superconductors \cite{Shi:2021nkx,Blencowe:2020ygo,Hirve:2024} and even quantum circuits \cite{Tokusumi:2018wii,Jannes:2010sa,Katayama:2021,Katayama:2022qmr} have one thing in common; an acoustic horizon. Recent experiments also utilize superfluids \cite{Jacobson:1998ms,Volovik:2005ga,Volovik:2006cz,Inui:2020,Braidotti:2021nhw,Jacquet:2022vak}, such as superfluid Helium. In some particular superfluid-Helium cases, the acoustic horizon can even be absent, i.e. there is no radial inflow of surface excitations. Recently, such laboratory contraptions have been constructed, dubbed \emph{multiply giant quantum vortex} \cite{Svancara:2023yrf,Smaniotto:2025hqm,Guerrero:2025kdn}, due to the enormous amount of vortex quanta used to build the vortex \cite{Alperin:2020aeq}. It has been shown that such experiments, when perturbed with surface waves, produce both QNMs and QBSs, regardless of the absence of an event horizon \cite{Svancara:2023yrf,Smaniotto:2025hqm,Guerrero:2025kdn}. Utilizing such a circulating vortex without radial flow, and calculating its spectrum is, nonetheless, absent from the current literature.

In this work we revisit an effective metric that describes acoustic excitations on a rotating geometry, with a tuning parameter that can turn the radial inflow on and off. This geometry can describe both the \emph{vortex flow} experiments involving superfluid Helium or quantum polaritonic fluids of light without radial flow \cite{Svancara:2023yrf,Smaniotto:2025hqm,Guerrero:2025kdn}, and rotating acoustic BH analogs with radial flow. Here, we focus on the case of a horizonless acoustic analog, where radial flow is absent, in an attempt to qualitatively replicate the results presented in \cite{Svancara:2023yrf}, regarding the QBSs. We find both the co-rotating and counter-rotating QBSs of the rotational horizonless vortex. These are qualitatively compared with the results in \cite{Svancara:2023yrf}. Our analysis matches well the qualitative results for co-rotating and counter-rotating QBSs, thus the theoretical setup used here can help into further elucidate the perturbative phenomenology of rotational horizonless vortices, which otherwise are experimentally sensitive to nonlinear dispersion and mechanical noise \cite{Svancara:2023yrf,Smaniotto:2025hqm,Guerrero:2025kdn}.

\section{Rotating acoustic tunable analogs}\label{RABHs}

We consider the acoustic BH solution in Minkowski spacetime obtained by Unruh \cite{Unruh:1980cg}. Then, we discuss a choice for the fluid flow velocity in order to obtain a rotating acoustic BH \cite{Visser:1997ux,Barcelo:2005fc} in cylindrical coordinates. By switching off the tuning parameter related to the radial inflow, we obtain an acoustic metric describing a purely rotational fluid flow. The fundamental equations of motion for an irrotational fluid are given by
\begin{eqnarray}
\nabla \times \mathbf{v} & = & 0, \label{eq:irrotational}\\
\partial_{t}\rho+\nabla\cdot(\rho\mathbf{v}) & = & 0, \label{eq:Continuity}\\
\rho[\partial_{t}\mathbf{v}+(\mathbf{v}\cdot\nabla)\mathbf{v}]+\nabla p & = & 0, \label{eq:Euler}
\end{eqnarray}
where $\mathbf{v}$, $\rho$, and $p$ are the velocity, density, and pressure of the fluid, respectively. We introduce the velocity potential $\Psi$, such that $\mathbf{v}=-\nabla\Psi$, and assume the fluid as barotropic, which means that $\rho=\rho(p)$. Then, by linearizing these equations of motion around some background $(\rho_{0},p_{0},\Psi_{0})$, namely,
\begin{eqnarray}
\rho & = & \rho_{0}+\epsilon\rho_{1}, \label{eq:rho}\\
p & = & p_{0}+\epsilon p_{1}, \label{eq:p}\\
\Psi & = & \Psi_{0}+\epsilon\Psi_{1}, \label{eq:psi}
\label{eq2:Madelung_representation}
\end{eqnarray}
we obtain the wave equation
\begin{align}
&-\partial_{t}\biggl[\frac{\partial \rho}{\partial p}\rho_{0}(\partial_{t}\Psi_{1}+\mathbf{v}_{0}\cdot\nabla\Psi_{1})\biggr]\nonumber
\\&+\nabla\cdot\biggl[\rho_{0}\nabla\Psi_{1}-\frac{\partial \rho}{\partial p}\rho_{0}\mathbf{v}_{0}(\partial_{t}\Psi_{1}+\mathbf{v}_{0}\cdot\nabla\Psi_{1})\biggr]=0,
\label{eq:wave_equation_Visser}
\end{align}
where the local speed of sound, $c_{s}$, is defined by
\begin{equation}
c_{s}^{-2} \equiv \frac{\partial \rho}{\partial p}.
\label{eq:sound}
\end{equation}
The wave equation (\ref{eq:wave_equation_Visser}) describes the propagation of the linearized scalar potential $\Psi_{1}$; that is, \emph{it governs the propagation of the phase fluctuations as weak excitations in a fluid's surface} 
that can be rewritten as a curved-spacetime wave equation
\begin{equation}
\frac{1}{\sqrt{-g}}\partial_{\mu}(g^{\mu\nu}\sqrt{-g}\partial_{\nu}\Psi_{1})=0.
\label{eq:KG_equation}
\end{equation}
where the effective curved spacetime $g_{\mu\nu}$ has a line element of the form
\begin{equation}
ds^{2} 
= \frac{\rho_{0}}{c_{s}}\left[-c_{s}^{2}dt^{2}+(dx^{i}-v_{0}^{i}dt)\delta_{ij}(dx^{j}-v_{0}^{j}dt)\right].
\label{eq:acoustic_metric}
\end{equation}
Unruh assumed the background flow as a spherically-symmetric, stationary inviscid fluid. By including circulation to the fluid flow, the acoustic metric given by Eq.~(\ref{eq:acoustic_metric}) obtains an explicit form in cylindrical coordinates, such as
\begin{align}
ds^{2}=&\frac{\rho_{0}}{c_{s}}\biggl\{-[c_{s}^{2}-(v_{0}^{r})^{2}-(v_{0}^{\theta})^{2}]dt^{2}+\frac{c_{s}^{2}}{c_{s}^{2}-(v_{0}^{r})^{2}}dr^{2}
\nonumber\\
&+r^{2}d\theta^{2}+dz^{2}-2v_{0}^{\theta}rdtd\theta\biggr\},
\label{eq:Unruh_circulation}
\end{align}
where the following coordinate transformations were performed
\begin{eqnarray}
&& dt \rightarrow dt-\frac{v_{0}^{r}}{c_{s}^{2}-(v_{0}^{r})^{2}}dr,\\
&& d\theta \rightarrow d\theta-\frac{v_{0}^{\theta}v_{0}^{r}}{r[c_{s}^{2}-(v_{0}^{r})^{2}]}dr,
\label{eq:coordinate_transformations}
\end{eqnarray}
with $\vec{v}=v_{0}^{r}\hat{r}+v_{0}^{\theta}\hat{\theta}$. Since we included rotation, the equation $c_{s}^{2}-(v_{0}^{r})^{2}-(v_{0}^{\theta})^{2}=0$ provides a condition for the effective ergosphere radius, which in turn is connected with energy extraction from BHs, and superradiance \cite{Brito:2015oca}. The radial and angular components of the fluid flow velocity can be written as
\begin{eqnarray}
&& v_{0}^{r}=-\frac{D}{r},\label{eq:radial_velocity}\\
&& v_{0}^{\theta}=\frac{C}{r}\label{eq:angular_velocity},
\end{eqnarray}
where $D$ and $C$ can be identified as the radial draining inflow parameter and the circulation, respectively. Therefore, we can write the line element for rotating acoustic BHs as
\begin{align}
\nonumber
ds^{2}=&-\biggl[f(r)-\frac{C^{2}}{r^{2}}\biggr]dt^{2}+\frac{1}{f(r)}dr^{2}\\
&+r^{2}d\theta^{2}+dz^{2}-2Cdtd\theta,
\label{eq:metric_RABHs}
\end{align}
where the acoustic lapse function $f(r)$ is given by
\begin{equation}
f(r)=1-\frac{D^{2}}{r^{2}}.
\label{eq:metric_function_RABHs}
\end{equation}
Note that we have dropped a position-independent prefactor by fixing $c_{s}=1$ and assumed a constant fluid density $\rho_{0}$. The ergosphere of this background can be found by solving
\begin{equation}
1-\frac{D^{2}}{r^{2}}=\frac{C^{2}}{r^{2}},
\label{eq:ergosphere_RABHs}
\end{equation}
which has a unique real solution given by
\begin{equation}
r_{\rm erg}=\sqrt{D^{2}+C^{2}}.
\label{eq:ergosphere_radius_RABHs}
\end{equation}
The acoustic event horizons form once the radial component of the fluid flow velocity exceeds the speed of sound, that is, when
\begin{equation}
f(r)=0,
\label{eq:surface_RABHs}
\end{equation}
which has two real solution given by
\begin{equation}
r_{\pm}=\pm D.
\label{eq:event_horizons_RABHs}
\end{equation}
Thus, $r=r_{+}$ designates a future acoustic event horizon (acoustic BH), while $r=r_{-}$ depicts a past acoustic event horizon (acoustic white hole). In addition, the dragging angular velocity of the acoustic horizon, $\Omega$, is given by
\begin{equation}
\Omega(r) \equiv -\frac{g_{03}}{g_{33}} = \frac{C}{r^{2}},
\label{eq:dragging_RABHs}
\end{equation}
such that $\Omega_{+}=\Omega(r_{+})=C/r_{+}^{2}$ is the frame dragging at the acoustic event horizon $r=r_{+}$.

\section{A vortex without radial inflow}\label{VFF}

When one nullifies the drain, i.e. when $D=0$, then the vortex becomes purely rotational and radial fluid flow is absent. In this case, the background presents no acoustic event horizon, due to the absence of radial inflow, and hence it describes a horizonless vortex fluid flow; an effective geometry that has been recently constructed in the laboratory with the use of superfluid Helium \cite{Svancara:2023yrf,Smaniotto:2025hqm} and quantum fluids of light \cite{Guerrero:2025kdn}.

A $(2+1)$-dimensional metric describing such a vortex model can be written as
\begin{equation}
ds^{2}=-\biggl(1-\frac{C^{2}}{r^{2}}\biggr)dt^{2}+dr^{2}+r^{2}d\theta^{2}-2Cdtd\theta,
\label{eq:metric_VFF}
\end{equation}
where we have set $z=h_{0}=\mbox{constant}$. In what follows, we will analyze the propagation of scalar excitations at the region of the general vortex spacetime given by Eq.~(\ref{eq:metric_VFF}), with $D=0$, and calculate the bound spectra. We will adopt an approach that includes the transformation of the equation of motion into a modified Mathieu differential equation. We will then obtain the bound states (BSs) numerically, as characteristic values of the modified Mathieu functions, which are intrinsically connected with the standard Mathieu functions \cite{Erdelyi1955}. In fact, the BSs are also solutions to the six boundary value problems that the standard Mathieu differential equation allows, thus they are eigenvalues \cite{Mclachlan_book,Meixner_book,Brimacombe:2020}. Finally, we will attempt to comprehend if there is a set of eigenfunctions that are more ``physical'' and closer to the qualitative behavior of the giant quantum vortex experiment in Ref. \cite{Svancara:2023yrf}. 

\subsection{Analogy between the effective metric used and modern laboratory experiments}

The metric depicted by Eq. \eqref{eq:metric_RABHs} describes a rotating acoustic BH, with an acoustic horizon at $r=r_+=D$, where the rotational velocity of the fluid is imprinted in the circulation parameter $C$. The way that Eq. \eqref{eq:metric_RABHs} is set-up allows for two tunable configurations; the typical rotating acoustic BH ($D>0$) or a rotating horizonless vortex without an acoustic horizon ($D=0$), i.e., a vortex without any radial inflow of surface fluctuations towards its center. The latter case presents a perfect testbed for contemporary laboratory experiments with fluids, superfluids and lasers. 

In particular, the experimental setup that we will try to qualitatively replicate from Ref. \cite{Svancara:2023yrf} has a lot of similarities with the geometry described by Eq. \eqref{eq:metric_VFF}. 
The key feature of the experiment in \cite{Svancara:2023yrf} is the use of superfluid $^4$He, that supports topological defects, known as quantum vortices, that carry circulation quanta. Each circulation quantum forms an irrotational (zero-curl) flow field in its vicinity. Due to such discretization, a vortex of superfluid $^4$He can manifest itself only as a \emph{giant} multiply quantized vortex. Such vortices can mimic rigid-body rotation and altogether suggest that an extensive draining vortex of $^4$He is a feasible candidate for experimental simulations of a quantum-field theory in curved spacetime. 
The setup also features a propeller that gives continuous circulation that can be loosely identified with the parameter $C$ in Eq. \eqref{eq:metric_VFF}. 
The most significant aspect of the apparatus in \cite{Svancara:2023yrf} is the absence of radial flow of interface fluctuations. This means that the experiment includes pure circulation and no acoustic horizon, as it is the case for the metric in Eq. \eqref{eq:metric_VFF}.

The geometry used here assumes the hydrodynamic limit, which effectively cancels any potential nonlinearities that emerge from dispersion. Not taking into consideration dispersive, nonhydrodynamic, effects practically leads to Unruh's historical acoustic metric proposal \cite{Unruh:1980cg} that formed the onset of the interest in hydrodynamic analog models of gravity and the construction of laboratory experiments. Nevertheless, avoiding dispersion in laboratory experiments is extremely difficult since nonlinearities are always part of realistic analog experiments \cite{Unruh:1994je,Unruh:2004zk,Lahav:2009wx,Macher:2009nz,Finazzi:2012iu,Svancara:2023yrf,Smaniotto:2025hqm,Guerrero:2025kdn}. Even so, simplified BH analog metrics like the one used here, can competently simulate the basic features and dynamics that take place in realistic analog BH experiments. 
 
Our perturbative scheme involves acoustic, non-dispersive, waves which is quite different from the dispersive interface waves of \cite{Svancara:2023yrf}. This is due to the use of a superfluid. Our metric is independent of the fluid's state or in general the state of matter used to create the vortex \cite{Svancara:2023yrf,Smaniotto:2025hqm,Guerrero:2025kdn}. Therefore, it can be applied for other experiments as well, though here we will commence the study by testing the robustness of our analysis by comparing our results with those in \cite{Svancara:2023yrf}.

\section{The bound spectrum of a horizonless rotating vortex}

\subsection{Scalar wave equation}

We are interested in the basic features of the effective geometry given by Eq.~(\ref{eq:metric_VFF}), in particular the propagation of excitations on the surface of the effective fluid flow. In order to perform this analysis, we have to solve the wave equation (\ref{eq:KG_equation}). By substituting Eqs.~(\ref{eq:metric_VFF}) into Eq.~(\ref{eq:KG_equation}), we obtain the following scalar wave equation
\begin{align}\nonumber
&-\frac{\partial^{2} \Psi_{1}}{\partial t^{2}}+\frac{1}{r}\frac{\partial}{\partial r}\biggl(r\frac{\partial \Psi_{1}}{\partial r}\biggr)+\biggl(\frac{1}{r^{2}}-\frac{C^{2}}{r^{4}}\biggr)\frac{\partial^{2} \Psi_{1}}{\partial \theta^{2}}\\
&-\frac{2C}{r^{2}}\frac{\partial^{2} \Psi_{1}}{\partial t \partial \theta}=0,
\label{eq:wave_equation_VFF}
\end{align}
where the scalar wave function is such that $\Psi_{1}=\Psi_{1}(t,r,\theta)$. Next, since the line element given by Eq. (\ref{eq:metric_VFF}) resembles those of rotating spacetimes, we can use the following separation ansatz for the perturbation $\Psi_1$
\begin{equation}
\Psi_{1} \simeq \mbox{e}^{-i \omega t}R(r)\mbox{e}^{i m \theta},
\label{eq:ansatz_VFF}
\end{equation}
where $\omega$ is the frequency (energy, in the natural units), $R(r)$ is the radial function, and $m$ is the azimuthal number (a discrete integer running from $-\infty$ to $+\infty$). By substituting Eqs.~(\ref{eq:metric_VFF})-(\ref{eq:ansatz_VFF}) into Eq.~(\ref{eq:KG_equation}), we obtain a radial master equation given by
\begin{equation}
R''(r)+\frac{1}{r}R'(r)+\biggl(\omega^{2}-\frac{m^{2}+2Cm\omega}{r^{2}}+\frac{C^{2}m^{2}}{r^{4}}\biggr)R(r)=0.
\label{eq:radial_equation_VFF_BSs}
\end{equation}

\subsection{Bound spectra}

Equation \eqref{eq:radial_equation_VFF_BSs} can be re-casted into the form of a modified Mathieu differential equation \cite{abramowitz_64} in order to be solved for its eigenvalues and eigenfunctions. We define a new radial coordinate $x$ such that
\begin{equation}
r=\tau\mbox{e}^{x}, \qquad r \in (0,\infty), \qquad x \in (-\infty,\infty).
\label{eq:x_VFF}
\end{equation}
By substituting \eqref{eq:x_VFF} into \eqref{eq:radial_equation_VFF_BSs} we obtain
\begin{equation}
R''(x)-\biggl(m^{2}+2Cm\omega-\frac{C^{2}m^{2}\mbox{e}^{-2x}}{\tau^{2}}-\tau^{2}\omega^{2}\mbox{e}^{-2x}\biggr)R(x)=0,
\label{eq:radial_2_VFF_BSs}
\end{equation}
or, after some algebraic calculations
\begin{equation}
R''(x)-[m^{2}+2Cm\omega-2Cm\omega\cosh(2x)]R(x)=0,
\label{eq:radial_2_VFF_BSs}
\end{equation}
where we have defined
\begin{equation}
\tau=\sqrt{\frac{Cm}{\omega}}.
\label{eq:tau_VFF}
\end{equation}
Thus, this transforms Eq. \eqref{eq:radial_equation_VFF_BSs} to the following normal form
\begin{equation}
R''(x)-[a-2q\cosh(2x)]R(x)=0.
\label{eq:modified_Mathieu_VFF}
\end{equation}
This is the modified Mathieu equation \cite{Mclachlan_book,Meixner_book,Erdelyi1955}, whose general solution
is given by
\begin{align}\nonumber
R(x) &= C_{1}\  \mbox{MathieuC}(a,q,ix)\\ 
&+ C_{2}\  \mbox{MathieuS}(a,q,ix),
\label{eq:radial_Mathieu_VFF}
\end{align}
where $C_{1}$ and $C_{2}$ are constants to be determined, $\mbox{MathieuC}$ and $\mbox{MathieuS}$ are given in terms of the even and odd standard Mathieu functions, respectively, and the parameters $a$ and $q$ are given by
\begin{eqnarray}
a & = & m^{2}+2Cm\omega,\label{eq:radial_a_Mathieu_VFF}\\
q & = & Cm\omega.\label{eq:radial_q_Mathieu_VFF}
\end{eqnarray}
Note that the modified Mathieu equation (\ref{eq:modified_Mathieu_VFF}) (and its eigenfunctions) differs from the standard Mathieu equation (\ref{eq:Mathieu_VFF}) (and its eigenfunctions) only by the input parameter $ix$ in the modified Mathieu functions and $x$ in the standard Mathieu functions. Frequently, Eq.~\ref{eq:modified_Mathieu_VFF}) appears in conjunction with Eq.~(\ref{eq:Mathieu_VFF}), when $a$ has one of the characteristic values $a_{n}(q)$ or $b_{n}(q)$, discussed below. We will restrict ourselves to this case, and obtain the BS frequencies as eigenvalues of the radial solution given by Eq. \eqref{eq:radial_Mathieu_VFF}. For a short review on the Mathieu functions, see Appendix \ref{Mathieu}.

The boundary value problems related to the modified Mathieu functions will define the physical meaning of the solutions with respect to an ``idealized'' version of the laboratory experiment we are trying to qualitatively reproduce. In fact, since the modified Mathieu functions have real eigenvalues, the eigensolutions we will obtain from our idealized experiment will be BSs, instead of QBSs. Nevertheless, we will see that to a certain extent, these eigensolutions describe the qualitative aspects of the dynamics of the actual laboratory experiment in \cite{Svancara:2023yrf}.

From a purely mathematical perspective, we cannot exclude any of the six eigenvalues. Thus, on the solution for the radial part of the massless Klein-Gordon equation in a vortex rotational fluid flow, given by Eq.~(\ref{eq:radial_Mathieu_VFF}), we will impose all six pairs of boundary value conditions, so that the parameter $a$ becomes eigenvalues of one of the six modified Mathieu functions, i.e.,
\begin{eqnarray}
\textrm{(I)} && \qquad  a=b_{n},\ \mbox{for}\ \mbox{Se}_{n}(x);\label{eq:I_modified_Mathieu}\\
\textrm{(II)} && \qquad  a=a_{n},\ \mbox{for}\ \mbox{Ce}_{n}(x);\label{eq:II_modified_Mathieu}\\
\textrm{(III)} && \qquad  a=b_{2n+2},\ \mbox{for}\ \mbox{Gey}_{2n+2}(x);\label{eq:III_modified_Mathieu}\\
\textrm{(IV)} && \qquad  a=b_{2n+1},\ \mbox{for}\ \mbox{Gey}_{2n+1}(x);\label{eq:IV_modified_Mathieu}\\
\textrm{(V)} && \qquad  a=a_{2n+1},\ \mbox{for}\ \mbox{Fey}_{2n+1}(x);\label{eq:V_modified_Mathieu}\\
\textrm{(VI)} && \qquad  a=a_{2n},\ \mbox{for}\ \mbox{Fey}_{2n}(x).\label{eq:VI_modified_Mathieu}
\end{eqnarray}
Here, $\mbox{Se}_{n}(x)=-i\ \mbox{se}_{n}(ix)$ and $\mbox{Ce}_{n}(x)=\mbox{ce}_{n}(ix)$ are the modified Mathieu functions of the first kind. Furthermore, $\mbox{Gey}_{n}(x)$ and $\mbox{Fey}_{n}(x)$ are the modified Mathieu functions of the second kind, which are given in terms of products between Mathieu functions of the first kind and Bessel functions of the second kind \cite{Erdelyi1955}. In fact, the parameters $a$ and $q$ are functions of $m$, $C$, $\omega$ and now of $n$. Hence, from the boundary value conditions given by Eqs.~(\ref{eq:I_modified_Mathieu})-(\ref{eq:VI_modified_Mathieu}) we can find the eigenvalues $\omega^{(i)}_{mn}$, which are the set of BS frequencies with azimuthal number $m$ and overtone number $n$, where $i=(\textrm{I},\ldots,\textrm{VI})$ labels the solution that corresponds to one of the six respective boundary value conditions imposed. To calculate the BS frequencies we solve numerically Eqs.~(\ref{eq:I_modified_Mathieu})-(\ref{eq:VI_modified_Mathieu}).

\begin{figure*}[t]
\centering
\includegraphics[scale=1]{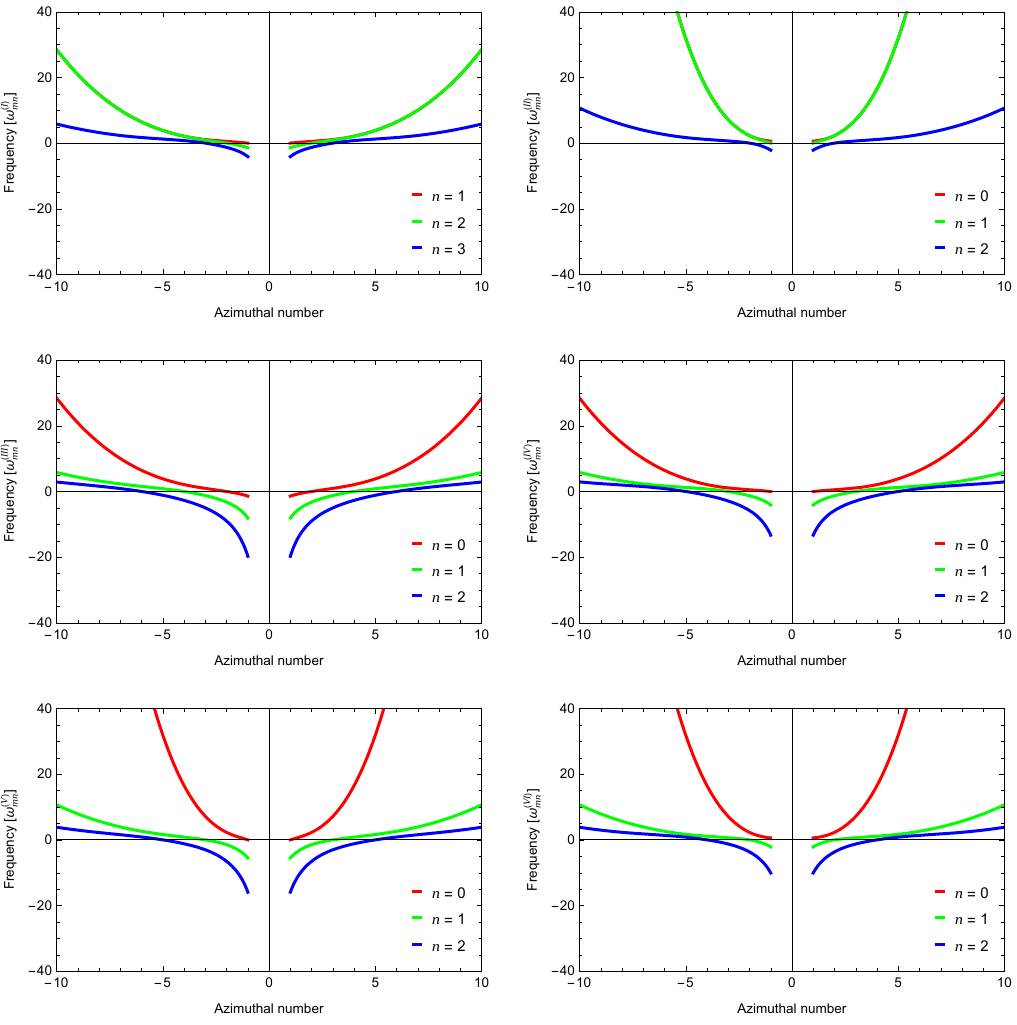}
\caption{Two-dimensional wave spectra for co-rotating and counter-rotating BSs for varying discrete azimuthal number $m$. We show the first three (non-zero) modes with unitary circulation $C=1$. Even though the subfigures depict continuous curves for the BSs, these are just the discrete BS frequencies for different $m$, that we joined together in a fit.}
\label{fig:Fig1_BSs_VFF}
\end{figure*}
\begin{figure*}[t]
\centering
\includegraphics[scale=1]{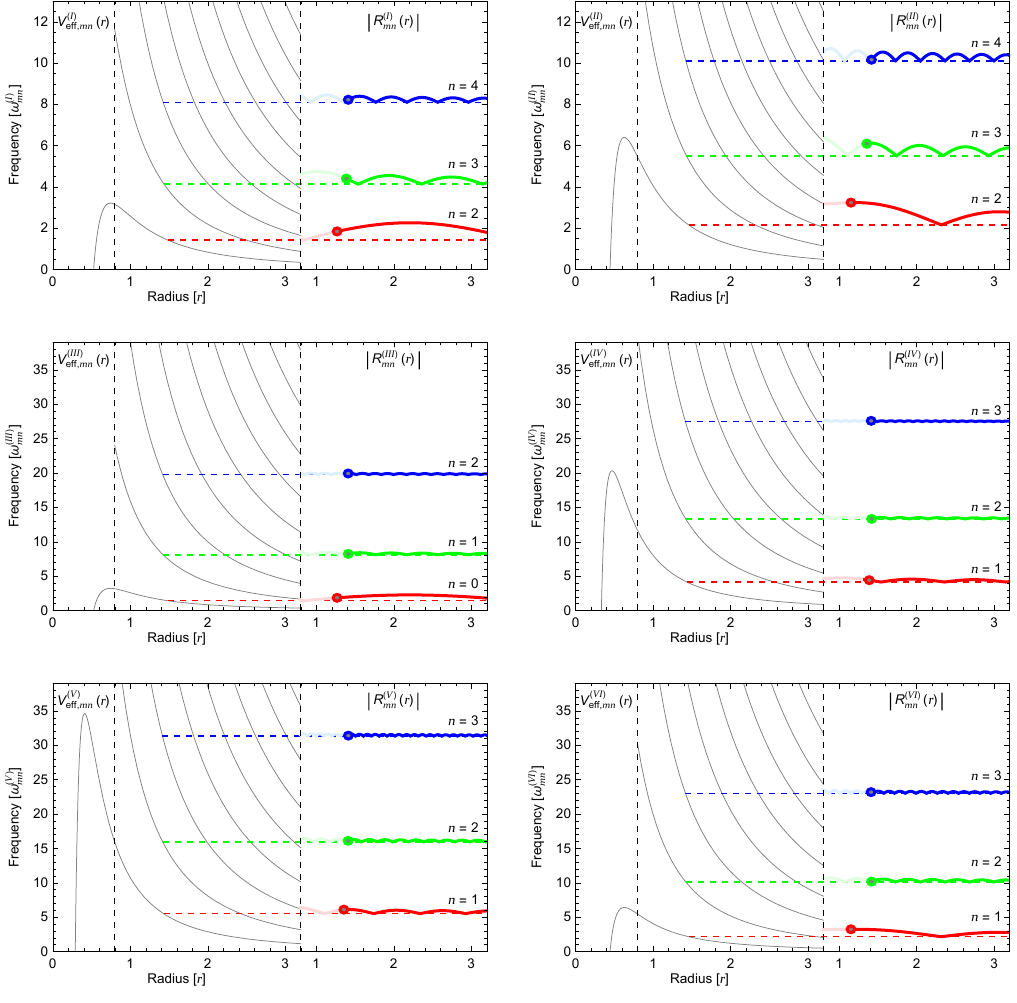}
\caption{Amplitudes of acoustic waves $|R^{(i)}_{mn}(r)|$ corresponding to the $m=1$ mode, with unitary circulation $C=1$. Here, we plot the first three (non-zero) overtones of the BSs, with respect to the radius $r$ ranging from 0 to $\pi$ and frequencies $\omega^{(i)}_{mn}$, for all possible boundary conditions $i=(\textrm{I},\ldots,\textrm{VI})$. We also show the effective potential $V^{(i)}_{{\rm eff},mn}$ with gray curves [the $\omega$-dependent factor, multiplying $R(r)$ in Eq. \eqref{eq:radial_equation_VFF_BSs}] for different BS frequency overtones. The solid colored lines correspond to the amplitude of rescaled BSs, which show a characteristic oscillatory pattern, that can be identified with their corresponding BS frequencies (dashed colored lines) outside the potential. The leftmost vertical dashed line corresponds to a reference point at $r=0.8$, while the rightmost one corresponds to $r=\pi$, as seen in the left subfigure of the effective potential, and $r=0.8$ as seen in the right subfigure of the wavefunction amplitudes. Crossing points with the effective potential are marked by gray points.} \label{fig:Fig2_BSs_VFF_modified}
\end{figure*}
\begin{figure*}[t]
\centering
\includegraphics[scale=1]{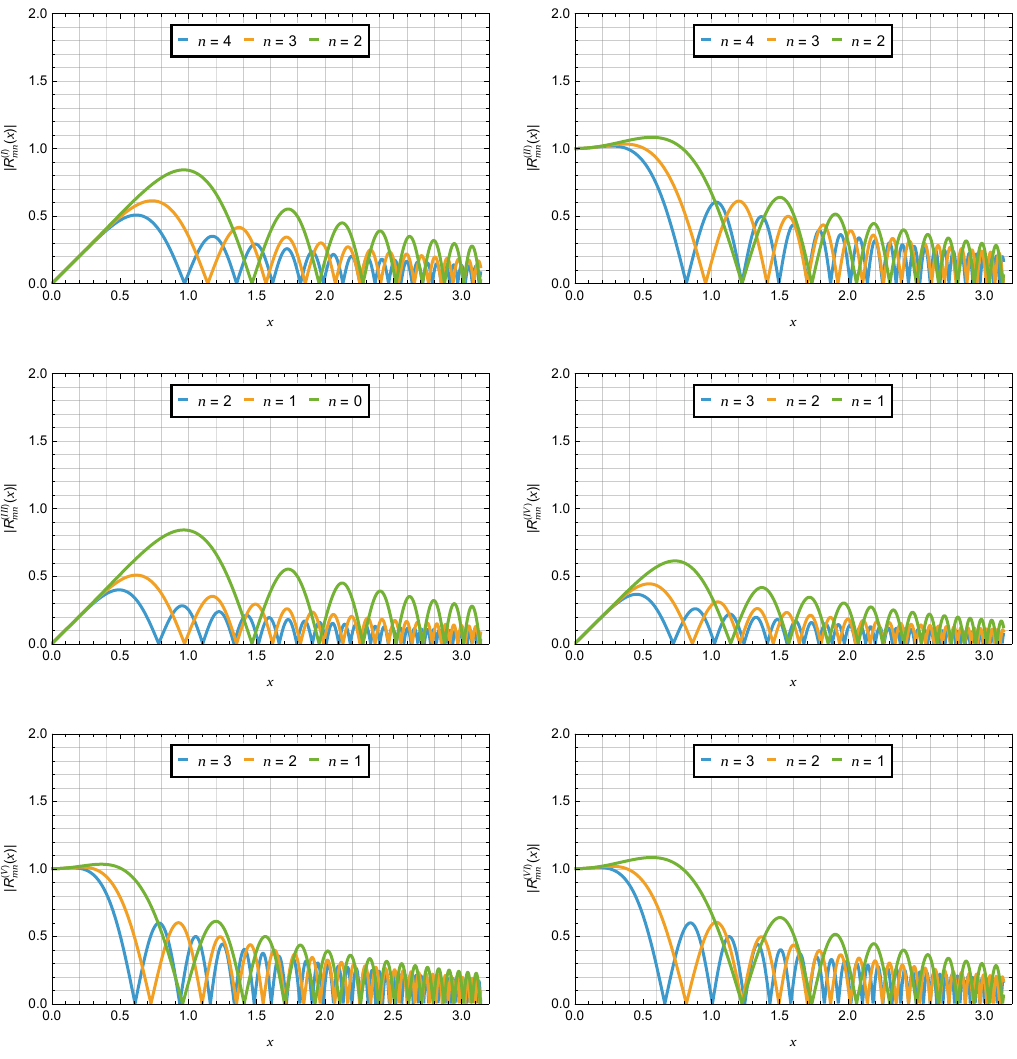}
\caption{Comparison between the absolute value of the first three (non-zero) wave eigenfunctions $|R^{(i)}_{mn}(x)|$, in the $x$-space of Mathieu functions, for all possible boundary conditions $i=(\textrm{I},\ldots,\textrm{VI})$. Here, the azimuthal number is $m=1$ and the circulation is $C=1$, where we vary the overtone number $n$.}\label{fig:Fig3_BSs_VFF_modified}
\end{figure*}

From a physical point of view, the boundary value problems imposed to the modified Mathieu equation \eqref{eq:radial_Mathieu_VFF} are pivotal to the resulting BS wavefunctions we are trying to reproduce. In the radial context of the problem, and knowing that the hyperbolic cosine will introduce blow-up behavior at $\pm\infty$, we need to apply boundary conditions that lead to regular (square-integrable) solutions. The modified Mathieu equation’s solutions grow or decay exponentially at large $|x|$. So, without constraints, one cannot get physically acceptable BSs. Typical boundary conditions for our problem should normalize the solutions of the modified Mathieu equation in order to obtain well-behaved BSs. These are quite straightforward, and can be found in many contexts such as vibrating systems, electromagnetism, elastic membranes and acoustics \cite{Mclachlan_book,Meixner_book}. For the problem in hand, we are studying fluctuations on a metric that describes a rotational acoustic analog without inflow. The first restriction comes from the absence of inflow of perturbations at the origin of the vortex $x=0$. We impose either a Dirichlet boundary condition, i.e., $R(0)=0$, or a Neumann boundary, i.e., $R^\prime(0)=0$ since solutions must remain finite or regular, there \cite{Mclachlan_book,Meixner_book}. At infinity, we can either impose purely outgoing waves, which lead to QBS solution that decay as they approach infinity, or impose, again, Dirichlet or Neumann boundary conditions, that lead to BSs. In accord to the conjugation between the modified and the standard Mathieu equations, we shall impose the only mathematically-allowed boundary conditions, shown in Appendix \ref{Mathieu}, Eqs. \eqref{eq:I_Mathieu}-\eqref{eq:VI_Mathieu}, that are connected with those in Eqs. \eqref{eq:I_modified_Mathieu}-\eqref{eq:VI_modified_Mathieu}, i.e. Dirichlet or Neumann depending on the case. These boundaries seems to satisfy the main principle of BS solutions and regularity\footnote{This is analogous to the choice of modified Bessel functions in classical radial problems, where one selects the decaying solutions, at the origin and outer boundary, for BSs.}.

Figure \ref{fig:Fig1_BSs_VFF} presents the theoretical predictions of the minimum frequency permissible for propagation of BSs for a given circulation $C$, and a variety of azimuthal numbers $m$ and overtones $n$. We consider both co-rotating ($m>0$) and counter-rotating ($m<0$) BS eigensolutions, from which the symmetry $\omega_{mn} \rightarrow -\omega_{-mn}$ arises. We note that this symmetry holds in our analysis due to the idealization of the experiment, that lacks nonlinear dispersion relations. This symmetry holds for all overtone number $n$. Notice that in some cases of boundary value conditions, the discrete BS frequency changes from positive to negative, though since they are purely real, the symmetry of the solution is still preserved. By comparing our resulting spectra with the experimental observations in \cite{Svancara:2023yrf} (see Figs. 2f and 2g), we find that their behavior for different boundary conditions remains qualitatively the same with those found in \cite{Svancara:2023yrf}, especially for radii of the superfluid interface that are much larger than the experimental potential barrier maximum. In fact the larger the radius, the more the aforementioned symmetry approximately holds in this particular experiment due to the reduction of the angular velocity distribution. Of course, in our case the choice of $C$ points to a velocity distribution, therefore when choosing $C$, we effectively change the radius of the interface, while ignoring other effects, such as dissipation of the wave's energy by viscous damping and by scattering into the vortex. Nevertheless, this is somewhat balanced by the stochastic drive originating from the fluid flow and mechanical vibrations in the experimental setup \cite{Svancara:2023yrf}, thus our assumptions and resulting spectral space $(m,f)$ of discrete BSs are not necessarily purely theoretical. Another similarity that appears between our data in Fig. \ref{fig:Fig1_BSs_VFF} and those in Fig. 2 of \cite{Svancara:2023yrf} is the fact that only a certain region of the spectral space is populated with BSs. We, also, observe that only some high-frequency (equivalent to high-energy) BSs have the capability to propagate.

Figure \ref{fig:Fig2_BSs_VFF_modified} shows the BS eigenfunctions with respect to the radial coordinate $r$. The demonstration unveils the pattern of the BSs with respect to the radius of the idealized vortex rotational fluid flow, for a fixed circulation $C=1$. The shapes of the effective potentials, for different overtone numbers $n$, are more pronounced closer to their respective peaks. Different boundary value problems give rise to different BS eigenfunctions, and respective BS eigenfrequencies, as well as higher- or lower-frequency maxima of the effective potential peaks. The most pronounced peak appears for the boundary boundary value problem (V). 

Interestingly, similar co-rotating BSs (dubbed QBSs) have been found in the laboratory experiment involving a giant quantum vortex of superfluid $^4$He \cite{Svancara:2023yrf}. Even though there is a resemblance between our results and those extracted from the realistic experiment, this is purely qualitative. Our effective spacetime metric does capture the important qualities and features of the realistic experiment, but it is not yet suitable to draw solid quantitative conclusions, especially because the patterns that appear in actual superfluid vortex experiments, as in \cite{Svancara:2023yrf}, regard interface waves, and not acoustic waves as the idealized manifestation of the effective metric and analysis involved. Our results, therefore, are qualitatively robust and may be of significance in the future when all aspects of the experimental setup are taken into account, such as nonlinear dispersion, interface waves, the transition between solid and hollow core vortices and, finally, which choice of boundary value problem is the most suitable one for such an experiment. 

For completeness, we have also analyzed the same problem with respect to the modified Mathieu-space radial coordinate $x$. Figure \ref{fig:Fig3_BSs_VFF_modified} demonstrates the absolute values of BS eigenfunctions for the six pairs of boundary value conditions with respect to the coordinate $x$. We focus on the co-rotating $m=1$ mode, with circulation set to $C=1$, though these choices are practically random since qualitatively similar results appear for higher $m$ and varying $C$. We observe that in the $x$-space, the boundary value conditions associated with the modified Mathieu equation, in conjunction with the standard Mathieu equation, that are imposed in order to obtain the BS eigensolutions, ensure boundedness and regularity. The cases in Fig. \ref{fig:Fig3_BSs_VFF_modified}, are divided into eigensolutions that start at $R(0)=0$, and those that start at $R(0)=1$. Their asymptotic behavior shows oscillatory decay in all cases. Thus, all wavefunctions are both regular at the core and bounded asymptotically. 

The most important question then is which boundary value condition is more physical for the calculation of BSs in their domain of existence. To find QBSs of BHs, the boundary conditions imposed are purely ingoing waves at the BH event horizon and purely reflecting waves at infinity \cite{Dolan:2007mj}. Here, the experiment we are trying to reproduce possesses no radial inflow of perturbations into the vortex, therefore there is no acoustic event horizon such as those appearing in BHs and BH analogs. Thus, a different boundary condition has to be imposed at the ``center'' of the vortex $x=0$. Since there is no event horizon, there is also an absence of a singularity at the internal vicinity of the vortex. This leads to the typical boundary condition that is imposed close to the center of stars when their perturbations are studied, i.e. regularity at the center of the star \cite{Andersson:1996pn,Allen:1997xj,Andersson:1999ks,Kokkotas:2000up,Gaertig:2008uz,Kruger:2019zuz}. The boundary condition at the center ensure that solutions remain finite and physically plausible. This is also the mathematical requirement to avoid singularities. 

To ensure regularity, the Neumann boundary condition specifies the values of the derivative applied at the boundary of a domain, which in our case may translates to $R^\prime(x=0) = \textrm{const.}$ in order for the eigensolutions to remain finite there. This means that there is no flow across the boundary, effectively making the central region a no-flux zone for the perturbation. In this context, it essentially implies that the perturbation is symmetric, as it occurs with solutions to the modified Mathieu equation which have certain characteristic eigenvalues. The normal derivative of the perturbed quantity (say the temperature or the density of the superfluid in \cite{Svancara:2023yrf}) is set to zero or a constant, meaning there is no change in the slope of the quantity across the internal region. 

The Dirichlet boundary on the other hand, where $R(x=0)=\textrm{const.}$ at the central region, ensures that there is no perturbation at the vicinity of the core whatsoever. Nevertheless, the constant resulting from the Neumann boundary condition can be chosen to be zero or otherwise, in order to cover wider scenarios of regularity, such as regularity at the inner (close to the central) region's boundary and the outer boundary of the rotational fluid that is deemed to possess only an azimuthal velocity distribution. From a mathematical perspective, the boundary value conditions that we can apply in modified Mathieu differential equations either set the field or its normal derivative equal to zero at $x=0$ [see Eqs. \eqref{eq:I_Mathieu}-\eqref{eq:VI_Mathieu}]. In principle, both a Dirichlet or a Neumann boundary condition can be used here, since they are both physically viable. Nevertheless, we could argue that in order to keep the temperature or the density of the superfluid constant at the central region of the vortex (and not zero, in accordance with Ref. \cite{Svancara:2023yrf}), then the most general boundary condition is the Neumann one that ensures both regularity and a non-zero finite constant at $x=0$. From our perspective, this narrows down the possible boundary value conditions of the modified Mathieu equation to \eqref{eq:II_Mathieu}, \eqref{eq:V_Mathieu} and \eqref{eq:VI_Mathieu}. 

As mentioned above, to find QBSs (or BSs) in BHs (or horizonless compact objects) we impose purely reflecting boundary conditions at infinity. This corresponds to Dirichlet boundary conditions at infinity, and in our case the eigenfunction should become zero at the asymptotic boundary of the modified Mathieu space that our solution exists. Following our argumentation, the pair of boundary value conditions needed in order for the resulting BSs to have a more appropriate physical meaning are the ones described by Eq.~(\ref{eq:V_modified_Mathieu}). This coincides with the boundary value condition (V) that allows for the highest frequency peak of the effective potential, which coincidentally looks more similar to what was found in the experimental apparatus of Ref. \cite{Svancara:2023yrf}. This, of course, is not a statement of quantitative agreement between the metric's perturbative behavior and the experimental data, but rather a mere coincidence. In principle, all boundary value conditions of the modified Mathieu equation can qualitatively reproduce the shapes and overall structure shown in Figs. 2, 4, and 5 of \cite{Svancara:2023yrf}.

\section{Final remarks}\label{Conclusions}

We have proposed the use of an acoustic metric that describes a draining and rotational fluid flow, specified by the drain $D$ and the circulation $C$ parameters, in order to qualitatively reproduce recent patterns of experimental data. This acoustic spacetime has been constructed assuming the hydrodynamic limit, that discards any nonlinear dispersion relations between the velocity density profile and the perturbations that we study. There are an abundance of applications of this metric in order to mimic analog BH experiments. A special case of interest in this work is the giant quantum vortex, recently constructed in the laboratory \cite{Svancara:2023yrf,Smaniotto:2025hqm,Guerrero:2025kdn}. The BS analysis of acoustic waves shows a lot of qualitative similarities with the actual experiment and seems universal, since there is no radial flow of perturbations in the experiment and in the effective metric ($D=0$). By obtaining a wave equation for the propagation of acoustic waves in the analog spacetime, we find a transformation that turns it into a modified Mathieu differential equation. Therefore, we find the solutions and frequencies through the modified Mathieu functions by solving all possible boundary value problems that these functions must, mathematically, satisfy. We then discuss their structure and compare them with the quasi-periodic interface BSs found in vortex experiments. Our findings indicate that the phenomenology of such an experiment could be qualitatively reproduced by our extremely simplified metric and perturbation analysis. We believe that our results for the BSs of acoustic waves might be helpful to the experimental community, since they can describe bounded acoustic waves propagating in a rotational fluid flow without radial inflow. In this sense, we expect that one could test the spacetime background under consideration and hence determine, in principle, if and how the idealized acoustic metric used here can provide more than just a qualitative picture.

To that end, we need to consider a key ingredient that is not directly discussed in the horizonless acoustic rotational analog. This is the ability to also tune the metric between a solid and a hollow core vortex, as it is performed in \cite{Svancara:2023yrf}, through particular choices of the propeller frequency. In order to be able to perform such a change to the metric, we need to include a new parameter and/or a functional of an existing parameter that enables us to perform this transition. The effective analog metric includes, besides the drain parameter $D$, a parameter for the circulation $C$ of the vortex. $C$ in turn, traces back to an angular velocity profile and a corresponding propeller frequency that controls the absence or presence of a hollow core. Thus, it might be the case that the circulation can be included in the metric in a form of a functional that enables this transition, depending on its value. This work needs to take advantage of the experimental data from Ref. \cite{Svancara:2023yrf}, in order to build (fit) a relation between various $C$ and the corresponding propeller frequencies of the experimental apparatus. We leave this work for the future.

Another interesting feature that can emerge from such experiments, and the resulting idealized metric under consideration, is the fact that there is no inflow of radial perturbations into the core. In a sense, the experiment may serve as a proxy for astrophysical environments; either extremely massive ones that surround galaxies, where the inflow of matter is negligible with respect to the lifetime of the environment (such as dark matter halos), or with dispersion relations that describe high-energy trans-Planckian effects around BHs. Hence, our analysis, and the successful control of the experiment in \cite{Svancara:2023yrf}, may lead the way for a more simplistic, yet accurate enough, design of decoupling environmental effects (matter perturbations) from the gravitational sector (especially polar gravitational perturbations in astrophysical BHs), and in general the relativistic treatment of astrophysical environments around BHs \cite{Barausse:2014tra,Jaramillo:2020tuu,Jaramillo:2021tmt,Cardoso:2021wlq,Destounis:2021lum,Cheung:2021bol,Speri:2022upm,Boyanov:2022ark,Courty:2023rxk,Cardoso:2022whc,Boyanov:2022ark,Destounis:2022obl,Vieira:2021ozg,Boyanov:2023qqf,Destounis:2023nmb,Cownden:2023dam,Arean:2023ejh,Sarkar:2023rhp,Eleni:2024fgs,Boyanov:2024fgc,Rosato:2024arw,Mollicone:2024lxy,Figueiredo:2023gas,Speeney:2024mas,Pezzella:2024tkf,Cardoso:2024mrw,Spieksma:2024voy,Cai:2025irl,Kouniatalis:2025itj,Fernandes:2025osu,Destounis:2025tjn}.

\begin{acknowledgments}
The authors are indebted to the anonymous referee for providing extremely insightful comments that led to a much more polished and mathematically sound version of the manuscript.
The authors are, also, indebted to Prof. Kostas D. Kokkotas for fruitful discussions and comments in the manuscript. The authors also would like to thank Pietro Smaniotto and Leonardo Solidoro for their help in better understanding the experimental apparatus and its function.
This study was financed in part by the Conselho Nacional de Desenvolvimento Científico e Tecnológico -- Brasil (CNPq) -- Research Project No. 440846/2023-4 and Research Fellowship No. 201221/2024-1. H.S.V. is partially supported by the Alexander von Humboldt-Stiftung/Foundation (Grant No. 1209836). Funded by the Federal Ministry of Education and Research (BMBF) and the Baden-W\"{u}rttemberg Ministry of Science as part of the Excellence Strategy of the German Federal and State Governments -- Reference No. 1.-31.3.2/0086017037.
K.D. acknowledges financial support provided by FCT – Fundação para a Ciência e a Tecnologia, I.P., under the Scientific Employment Stimulus – Individual Call – Grant No. 2023.07417.CEECIND/CP2830/CT0008.
This project has received funding from the European Union’s Horizon-MSCA-2022 research and innovation programme ``Einstein Waves'' under grant agreement No. 101131233.
\end{acknowledgments}

\appendix

\section{Standard Mathieu equation and its periodic solutions}\label{Mathieu}

The standard form of the Mathieu equation is given by \cite{Erdelyi1955}
\begin{equation}
R''(x)+[a-2q\cos(2x)]R(x)=0,
\label{eq:Mathieu_VFF}
\end{equation}
whose general solution is given by
\begin{align}\nonumber
R(x) &= C_{1}\  \mbox{MathieuC}(a,q,x) \\
&+ C_{2}\  \mbox{MathieuS}(a,q,x),
\label{eq:general_Mathieu_VFF}
\end{align}
where $C_{1}$ and $C_{2}$ are constants to be determined, $\mbox{MathieuC}$ and $\mbox{MathieuS}$ are the even and odd standard Mathieu functions, respectively. The Mathieu functions become periodic for some characteristic values of $a$ for given $q$. In any half-open interval of length $\pi$ on the real axis, an even Mathieu function of the first kind with $n$ zeros in the interval $0 \le x < \pi$ is denoted by $\mbox{ce}_{n}(x,q)$, and has characteristic values $a_{n}(q)$, with $n=0,1,2,\ldots$, while an odd Mathieu function of the first kind is denoted by $\mbox{se}_{n}(x,q)$, and has characteristic values $b_{n}(q)$, with $n=1,2,\ldots$. From now on, we will write $\mbox{ce}_{n}(x)$, $\mbox{se}_{n}(x)$, $a_{n}$ and $b_{n}$, by omitting $q$. These characteristic values satisfy any of the following pairs of boundary conditions \cite{Erdelyi1955}
\begin{widetext}
\begin{eqnarray}
\textrm{(I)} && \qquad  R(0)=R(\pi)=0,\ \mbox{for}\ \mbox{se}_{n}(x)\ \mbox{and}\ b_{n},\ \mbox{with period of}\ \pi;\label{eq:I_Mathieu}\\
\textrm{(II)} && \qquad  R'(0)=R'(\pi)=0,\ \mbox{for}\ \mbox{ce}_{n}(x)\ \mbox{and}\ a_{n},\ \mbox{with period of}\ \pi\label{eq:II_Mathieu}.
\end{eqnarray}
\end{widetext}
On the other hand, if $R(x)$ is either $\mbox{se}_{n}(x)$ or $\mbox{ce}_{n}(x)$, then $R(x)$ and $R(\pi-x)$ satisfy the same differential equation and the same boundary conditions, and must be constant multiples of each other so that $R(x)$ is either an even or an odd function of $\pi/2-x$. This leads to four additional pairs of boundary conditions \cite{Erdelyi1955}:
\begin{widetext}
\begin{eqnarray}
\textrm{(III)} && \qquad  R(0)=R(\pi/2)=0,\ \mbox{for}\ \mbox{se}_{2n+2}(x)\ \mbox{and}\ b_{2n+2},\ \mbox{with period of}\ \pi;\label{eq:III_Mathieu}\\
\textrm{(IV)} && \qquad  R(0)=R'(\pi/2)=0,\ \mbox{for}\ \mbox{se}_{2n+1}(x)\ \mbox{and}\ b_{2n+1},\ \mbox{with period of}\ 2\pi;\label{eq:IV_Mathieu}\\
\textrm{(V)} && \qquad  R'(0)=R(\pi/2)=0,\ \mbox{for}\ \mbox{ce}_{2n+1}(x)\ \mbox{and}\ a_{2n+1},\ \mbox{with period of}\ 2\pi;\label{eq:V_Mathieu}\\
\textrm{(VI)} && \qquad  R'(0)=R'(\pi/2)=0,\ \mbox{for}\ \mbox{ce}_{2n}(x)\ \mbox{and}\ a_{2n},\ \mbox{with period of}\ \pi\label{eq:VI_Mathieu}.
\end{eqnarray}
\end{widetext}
Here, for each $n=0,1,2,\ldots$ there is exactly one characteristic function of each of these four boundary value problems, and $n$ is the number of zeros in the interval $0 < x < \pi/2$.

\bibliography{biblio}

\end{document}